\documentclass[notitlepage,nofootinbib,preprintnumbers,amssymb,superscriptaddress]{revtex4-1}
\usepackage{amsfonts,amssymb,amsmath,graphicx,color,bm}
\definecolor{ultramarine}{rgb}{0.07, 0.04, 0.56}
\definecolor{cadmiumgreen}{rgb}{0.0, 0.42, 0.24}
\definecolor{indigo(dye)}{rgb}{0.0, 0.25, 0.42}
\usepackage[linktocpage=true]{hyperref}
\hypersetup{
colorlinks=true,
citecolor=green,
linkcolor=cadmiumgreen,
urlcolor=indigo(dye),
}

\usepackage{autobreak}

\def\[{\begin{equation}}
\def\]{\end{equation}}

\newcommand*{\cH}{{\mathcal{H}}}
\newcommand*{\cF}{{\mathcal{F}}}
\newcommand*{\cG}{{\mathcal{G}}}

\newcommand*{\cJ}{{\mathcal{J}}}
\newcommand*{\cV}{{\mathcal{V}}}
\newcommand*{\cZ}{{\mathcal{Z}}}
\newcommand*{\cK}{{\mathcal{K}}}
\newcommand*{\cQ}{{\mathcal{Q}}}
\newcommand*{\cM}{{\mathcal{M}}}

\begin{document}

\title{Time-dependent, spherically symmetric background in Kaluza-Klein compactified Horndeski theory and the speed of gravity waves}

\author{S. Mironov}
\email{sa.mironov\_1@physics.msu.ru}
\affiliation{Institute for Nuclear Research of the Russian Academy of Sciences,
60th October Anniversary Prospect, 7a, 117312 Moscow, Russia}
\affiliation{Institute for Theoretical and Mathematical Physics,
MSU, 119991 Moscow, Russia}
\affiliation{NRC, "Kurchatov Institute", 123182, Moscow, Russia}

\author{M. Sharov}
\email{sharov.mr22@physics.msu.ru}
\affiliation{Institute for Nuclear Research of the Russian Academy of Sciences,
60th October Anniversary Prospect, 7a, 117312 Moscow, Russia}
\affiliation{Department of Particle Physics and Cosmology, Physics Faculty, M.V. Lomonosov Moscow State University,
Vorobjevy Gory, 119991 Moscow, Russia}

\author{V. Volkova}
\email{volkova.viktoriya@physics.msu.ru}
\affiliation{Institute for Nuclear Research of the Russian Academy of Sciences,
60th October Anniversary Prospect, 7a, 117312 Moscow, Russia}
\affiliation{Department of Particle Physics and Cosmology, Physics Faculty, M.V. Lomonosov Moscow State University,
Vorobjevy Gory, 119991 Moscow, Russia}

\begin{abstract}
We revisit the models recently derived from a Kaluza-Klein compactification of higher dimensional Horndeski theory, where the resulting electromagnetic sector features non-trivial couplings to Horndeski scalar. In particular, this class of theories admits the electromagnetic waves propagating at non-unit speed, which in turn allows to relax the constraints on Horndeski theories following from multi-messenger speed test. In this work we prove that both gravitational wave and its electromagnetic counterpart propagate at the same, although non-unit, speed 
above an arbitrarily time-dependent, spherically symmetric background within the theories in question. Hence, we support the statement that several subclasses of Horndeski theories are not necessarily ruled out after the GW170817 event provided the photon-Galileon couplings are allowed. We also formulate the set stability conditions for an arbitraty solution within the discussed theoretical setting.
\end{abstract}

\maketitle

\section{Introduction}\label{sec:intro}

Multi-messenger gravitational wave (GW) astronomy with the detection of
GW170817 event~\cite{LIGOScientific:2017vwq} and its electromagnetic (EM) counterpart GRB170817A~\cite{Goldstein:2017mmi} 
made it possible to constrain the speed difference of GWs and light down to 
$|{c_{GW}}/{c_{\gamma}}-1| \leq 5 \times 10^{-16}$~\cite{LIGOScientific:2017zic}.  
The latter drastically restricts possible modifications to General Relativity in the late-time Universe, since modifying gravity in many cases results in the  "anomalous" propagation speed of GW which significantly differs from that of light. 
In particular, Horndeski theories~\cite{Horndeski:1974wa,Deffayet:2011gz,Kobayashi:2011nu}
which generally
represent a wide class of healthy scalar-tensor theories, have faced considerable limitations in order to reproduce the equality $c_{GW}=c_{\gamma}$
\footnote{Here and in what follows for theoretical simplicity we consider strict equality of the speeds, while the experimental constraints still allow for minor deviations.},
so that only a few relatively simple subclasses have survived within a homogeneous Universe setting as Dark Energy candidates~\cite{Ezquiaga:2017ekz,Creminelli:2017sry,Baker:2017hug,Langlois:2017dyl,Kase:2018aps} (see, however,~\cite{deRham:2018red}).

These constraints from multi-messenger speed test, however, apply only to Horndeski theory in the case when the photon is minimally coupled to gravity and $c_{\gamma} = 1$. 
It has been recently shown in~\cite{KK,KK2} that some of the constraints from GW170817 on the Lagrangian functions in Horndeski theory get relaxed, if one allows non-trivial couplings of the photon and the scalar field similar in form to those for the graviton and the scalar field already present in the Horndeski Lagrangian. Technically this type of scalar-vector-tensor theories arises in a natural way in result of Kaluza-Klein compactification of the 5D Horndeski theory on a circle. 
In result, the speed of gravitational and electromagnetic waves get modified in the same way so that the equality $c_{GW} = c_{\gamma}$ holds automatically in such scalar-vector-tensor theories. Even though both speeds are generally non-unit, i.e. $c_{GW} = c_{\gamma} \neq 1$, their ratio remains constant 
for a wider subclass of Horndeski theories than originally allowed by GW170817 constraints, e.g. with non-trivial $G_5$ (see eq.~\eqref{eq:G5_5DGalileonsIn4D} in Sec.~\ref{sec:setup}). 
Recently the concept of non-minimal dark energy-photon
coupling  in the context of multi-messenger speed test has been further developed within beyond Horndeski theory in~\cite{Babichev:2024kfo}.

While the calculations in~\cite{KK,KK2} have verified the equality relation 
$c_{GW} = c_{\gamma}$
in a cosmological setting,
in this note we generalize the result and check whether the speeds' equality holds in significantly curved backgrounds, where the 
scalar-field gradients are non-vanishing. 
To do so in Sec.~\ref{sec:setup} within the class of scalar-vector-tensor theories discussed in~\cite{KK} we consider a spherically symmetric background, where all fields that are involved feature arbitrary time-dependence.
We derive
the quadratic action for {spin 2 and spin 1} axial (or parity odd) linear perturbations 
over the chosen background
in full generality in Sec.~\ref{sec:action} and 
prove in Sec.~\ref{sec:speeds}
that the propagations speeds of both GW and the electromagnetic wave are generally indentical in this dynamical inhomogeneous setting. 
In Sec.~\ref{sec:stability} we formulate a set of stability constraints, which ensures the absence of ghosts and gradient instabilities among the high energy modes.  
We briefly conclude and discuss the results in Sec.~\ref{sec:conclusion}.

\section{Compactified 5D Horndeski into luminal scalar-vector-tensor theory}
\label{sec:setup}

The authors of Ref.~\cite{KK} first consider quadratic Horndeski theory
\footnote{We adopt the naming for Horndeski theory subclasses which is based on the highest power of the scalar field $\pi$ involved. }
 with non-trivial functions $G_2$, $G_3$ and $G_4$ in $D=5$ and compactify it on a circle. 
The corresponding Lagrangian after compactification reads (see~\cite{KK2} for technical details):
\begin{multline}
\label{eq:G4_5DGalileonsIn4D}
\phi({\mathcal{L}_2+\mathcal{L}_3+\mathcal{L}_4})+\mathcal{L}_{4A_{\mu}}+\mathcal{L}_{4\phi} = 
\int \text{d}^4 x\,\sqrt{-g}\,\phi \, \left[
\underline{G_2(\pi,\,X)+G_3(\pi,\,X) \, \Box \pi} +
G_4(\pi,\,X)\, \left(\underline{ R}-\frac{1}{4}\phi^2\,F^2-2\frac{\Box \phi}{\phi}\right) 
\right.\\\left. 
+ G_{4,X}(\pi,\,X)\,\Big(\underline{(\Box \pi)^2-(\nabla_\mu\nabla_\nu \pi)^2} + 2\,\frac{1}{\phi}\, \nabla_\mu\phi \nabla^\mu \pi\,\Box\pi
-  \frac{1}{2}\phi^2\, F_\mu{}^\sigma\, F_{\nu\sigma}\,\nabla^\mu \pi\,\nabla^\nu \pi 
\Big)\right]\,,
\end{multline}
where $F_{\mu\nu} = \partial_{\mu} A_{\nu} - \partial_{\nu} A_{\mu}$, $F^2\equiv F^{\mu\nu}F_{\mu\nu} $, $G_{4X}=\partial G_4/\partial X$ and $X= -\frac12\: g^{\mu\nu}\pi_{,\mu}\pi_{,\nu}$. The resulting theory in $D=4$ involves the metric field $g_{\mu\nu}$, the scalar field $\pi$ of quadratic Horndeski type (described by $\mathcal{L}_2+\mathcal{L}_3+\mathcal{L}_4$, i.e. the underlined terms in eq.~\eqref{eq:G4_5DGalileonsIn4D}), 
the $U(1)$ gauge field $A_{\mu}$ and the dilaton $\phi$.
Note that the Lagrangian $\mathcal{L}_{4A_{\mu}}$ for the vector field $A_{\mu}$, which is identified with the electromagnetic field, 
involves non-trivial couplings to the scalar field $\pi$ (we refer to $\pi$ as Galileon field in what follows).
As it was explicitly shown in~\cite{KK}
these couplings are respondible for the required modification of photon's speed, so that $c_{GW} = c_{\gamma}$ for any choice of scalar potentials $G_2$, $G_3$ and $G_4$ in eq.~\eqref{eq:G4_5DGalileonsIn4D} in a Friedmann-Lemaitre-Robertson-Walker (FLRW) setting.

Similar compactification procedure was carried out in~\cite{KK,KK2} for a specific subclass of cubic Horndeski theory with $G_5(\pi,X) \equiv G_5(\pi)$. The resulting Lagrangian has the form:
\begin{eqnarray}
\label{eq:G5_5DGalileonsIn4D}
\phi\mathcal{L}_5+\mathcal{L}_{5A_{\mu}}+\mathcal{L}_{5\phi}
&=&\int \text{d}^4 x\,\sqrt{-g}\,\phi\,G_5(\pi)\, \Bigg[\left( \underline{R^{\mu\nu}-\frac{1}{2}g^{\mu\nu}\,R}\right) \nabla_\mu \nabla_\nu \pi  
-\frac{1}{2\,\phi}R\, {{\nabla_\mu \phi}\,\nabla^\mu \pi }
\\&&
+\frac{1}{\phi}\left(\Box\phi\,\Box\pi-\nabla_\mu\nabla_\nu \phi \nabla^\mu\nabla^\nu \pi\right) +\frac{1}{2}\phi^2\,F_{\mu\nu}\,\nabla_\sigma F^{\nu\sigma}\,\nabla^\mu \pi  
\nonumber\\
&&
+ \frac{1}{8}\phi\, F^{\mu\nu}\,F^{\sigma\rho}\bigg(3\,g_{\nu\rho}(-4\,g_{\lambda\mu}\,g_{\beta\sigma}+g_{\lambda\beta}\,g_{\mu\sigma})\,\nabla^\lambda \pi \nabla^\beta \phi 
 +\phi\, g_{\sigma\mu}\,(-4\,\nabla_\nu\nabla_\rho \pi +g_{\rho\nu}\Box\pi) \bigg)
\Bigg]\nonumber, 
\end{eqnarray}
where $\mathcal{L}_5$ denotes the standard cubic Horndeski subclass 
with $G_5(\pi)$ (see the underlined terms on the right-hand side).
Again for any choice of the potential $G_5(\pi)$ the theory features "luminal" gravitational waves, while the speed of the photon is modified and is generally non-unit. Let us note that both Lagrangians~\eqref{eq:G4_5DGalileonsIn4D} and~\eqref{eq:G5_5DGalileonsIn4D} can be dubbed "vector-scalar Galileon theory", meaning that both the Galileon field and the electromagnetic field feature higher derivatives in the Lagrangian, while the equations of motion are still second order. 
{Let us note that the dilaton does not play a significant role in our reasoning since it does not modify neither $c_{GW}$ nor $c_{\gamma}$ as we will see in Sec.~\ref{sec:speeds}.}

The aim of this note is to check whether the equality of the
graviton and photon speeds in theories with the Lagrangians~\eqref{eq:G4_5DGalileonsIn4D} and~\eqref{eq:G5_5DGalileonsIn4D} holds above 
more general backgrounds than that of homogeneous cosmology discussed in~\cite{KK}. 
In what follows we 
consider a time-dependent, spherically symmetric background geometry
of the following form:
\[
\label{eq:backgr_metric}
ds^2 = - A(t,r)\:dt^2 + \frac{dr^2}{B(t,r)} + J^2(t,r) \left(d\theta^2 + \sin^2\theta\: d\varphi^2\right),
\]
where the metric variables are arbitrary functions of time and radial coordinates. The background Galileon, dilaton and electromagnetic field are also arbitrarily dynamical and inhomogeneous: 
\[
\label{eq:backgr_fields}
\pi=\pi(t,r), \qquad \phi=\phi(t,r), 
\qquad A_{\mu} = (A_{0}(t,r),\;  A_1(t,r),\; 0,\; 0).
\]
For this configuration of the scalar field and background metric, the kinetic terms reads:
$$X = \dfrac{\dot{\pi}^2}{2 A} - \dfrac12 B{\pi'}^2,$$
where dot and prime denote derivatives w.r.t. time and radial coordinate, respectively.

To derive and compare the propagations speeds for both gravity and electromagnetic waves we consider linear perturbations of metric and vector field $A_{\mu}$ above the background~\eqref{eq:backgr_metric}-\eqref{eq:backgr_fields}, which features both time-dependence and inhomogeneity.
In the following section we derive the corresponding quadratic action for both perturbation modes in theories~\eqref{eq:G4_5DGalileonsIn4D} and~\eqref{eq:G5_5DGalileonsIn4D}.
From the technical point of view the following procedure is highly similar to that carried out in~\cite{MSV} for pure (beyond) Horndeski theory (see also~\cite{Baez:2022rdz}) with the background metric of the form~\eqref{eq:backgr_metric}.

\section{Odd-parity perturbations}
\label{sec:odd_sector}

Let us consider linear perturbations 
within Regge-Wheeler formalism~\cite{ReggeWheeler}, where perturbations are divided into parity odd and parity even modes, according to their transformation under two-dimensional reflection. Since the theories~\eqref{eq:G4_5DGalileonsIn4D} and~\eqref{eq:G5_5DGalileonsIn4D} do not feature parity violating terms, the odd- and even-parity modes evolve independently. 
In what follows we focus on odd-parity sector since it contains the two necessary modes: one perturbation mode comes from the metric and the other one is sourced by the vector field. 
Hence, the analysis of parity odd sector is sufficient for derivation of the speeds for graviton and photon. 

Since neither Galileon ${\pi}$ nor dilaton ${\phi}$ fields contribute to the odd-parity sector, we consider perturbations of only
metric and vector fields
\[
\label{eq perturbations}
g_{\mu\nu} \to g_{\mu\nu} + h_{\mu\nu}, \qquad A_{\mu} \to A_{\mu} + \delta A_{\mu},
\]
which can be written in terms of spherical harmonics $Y_{\ell m}(\theta,\phi)$ as follows
\begin{eqnarray}
     &  & h_{tt}=0,~~~h_{tr}=0,~~~h_{rr}=0,\\
     &  & h_{ta}=\sum_{\ell, m} h_{0,\ell m}(t,r)E_{ab}\nabla^{b}Y_{\ell m}(\theta,\varphi),    \label{eq:pertV1} \\
     &  & h_{ra}=\sum_{\ell, m} h_{1,\ell m}(t,r)E_{ab}\nabla^{b}Y_{\ell m}(\theta,\varphi),    \label{eq:pertV2} \\
     &  & h_{ab}=\frac{1}{2}\sum_{\ell, m} h_{2,\ell m} (t,r)\left[E_{a}^{~c}\nabla_{c}\nabla_{b}Y_{\ell m}(\theta,\varphi)+E_{b}^{~c}\nabla_{c}\nabla_{a}Y_{\ell m}(\theta,\varphi)\right],
     \label{odd-ab}
\end{eqnarray}
and
\begin{align}
    \delta A_t = \delta A_r=0 \,, 
    \qquad 
    \delta A_a = \sum_{\ell m} \mathcal{V}_{\ell m}(t,r) E_{ab} \nabla^b Y_\ell^m(\theta,\varphi) \, ,
    \label{vectorper}
\end{align}
where $a,b = \theta,\varphi$, $E_{ab} = \sqrt{\det \gamma}\: \epsilon_{ab}$, with $\gamma_{ab} = \mbox{diag}(1, \;\sin^2\theta)$;
$\epsilon_{ab}$ is
totally antisymmetric symbol ($\epsilon_{\theta\varphi} = 1$) and $\nabla_a$
is covariant derivative on a 2-sphere. The modes with different $(\ell,m)$ evolve independently, hence, in our calculations we consider a specific mode and drop the indexes $\ell$ and $m$ for brevity. Moreover, to simplify things we take advantage of the spherical symmetry and set $m=0$ without loss of generality. 

In what follows we make use of the general covariance and by utilizing the infinitesimal coordinate transformations $x^{\mu}\to x^{\mu}+\xi^{\mu}$ we 
adopt the Regge-Wheeler gauge $h_2 = 0$ for modes with $\ell \geq 2$.

\subsection{Quadratic action for perturbations with $\ell \geq 2$}
\label{sec:action}

Following the standard approach, we substitute
the linearized metric~\eqref{eq:backgr_metric} into actions~\eqref{eq:G4_5DGalileonsIn4D} and~\eqref{eq:G5_5DGalileonsIn4D}, then expand
the result up to second order in perturbations and adopt the gauge choice $h_2=0$. Then
the quadratic action in the odd-parity sector reads:
\begin{multline}
\label{eq:action_odd_metric}
S^{(2)}_{h+\cV} = \int \mbox{d}t\:\mbox{d}r \left[ a_1 h_0^2+a_2 h_1^2+a_3 \left( {\dot h_1}^2-2 {\dot h_1} h_0'+{h_0'}^2+\frac{4 J'}{J} {\dot h_1} h_0 
+ \frac{4 \dot{J}}{J} h_1 h_0' \right) 
+  a_4 h_1 h_0 
\right.\\\left.
+ b_1 (\cV' h_0 - \dot{\cV} h_1 )
+ b_2 \dot{\cV}^2 
+ b_3 {\cV'}^2
+ b_4 \dot{\cV} \cV' 
+ b_5 \cV^2\right],
\end{multline}
where we have integrated over angles and used the background equations of motion to simplify $a_i$ coefficients
\footnote{This is the only place throughout the paper where we make use of equations of motion following from~\eqref{eq:G4_5DGalileonsIn4D} and~\eqref{eq:G5_5DGalileonsIn4D}. Due to the bulky form of the equations and their relatively low significance for our key argument we do not present them explicitly.}
, which explicitly read:
\[
\label{eq:A1A2A3}
\begin{aligned}
& a_1 = \frac{\ell (\ell+1)}{J^2}
\left[\frac{d}{d r}\left(
J\: J' \sqrt{\frac{B}{A}}{\cal H}\right)
+\frac{(\ell-1)(\ell+2)}{2\sqrt{AB}}{\cF}
\right],  
\qquad\qquad 
 b_1 = \ell(\ell+1) \sqrt{\frac{B}{A}}  \phi^2 (\dot{A}_1-A_0') \cH, &
\\
& a_2 =  \frac{\ell (\ell+1)}{J^2} \left[ \frac{d}{d t}\left(
J\: \dot{J} \sqrt{\frac{B}{A}}{\cal H}\right) - {(\ell-1) (\ell+2)}\frac{\sqrt{AB} }{2}{\cG}\right],
\qquad
 b_2 = \frac{\ell (\ell+1)}{2}  \frac{\phi^2}{\sqrt{AB}} \cF, & 
\\
& a_3 = \dfrac{\ell(\ell+1)}{2} \sqrt{\frac{B}{A}} {\mathcal{H}},
\qquad\qquad\qquad\qquad\qquad\qquad\qquad\qquad\qquad ~
b_3 = - \frac{\ell (\ell+1)}{2} \phi^2\sqrt{AB} \cG ,&
\\
& a_4 = \frac{\ell(\ell+1)}{J^2}\sqrt{\frac{B}{A}} 
\left[(\ell-1)(\ell+2)  \cJ - 4 \cH J' \dot{J}\right], 
\qquad\qquad\qquad ~~~
b_4 = \ell (\ell+1) \sqrt{\frac{B}{A}}  \phi^2 \cJ, &
\end{aligned} 
\]
and
\begin{subequations}
\label{eq:HGF}
\begin{align}
& \cF = 2 \phi \left[G_4 - G_{4X} \frac{\dot{\pi}^2}{ A} - G_{5\pi}\left(X - \frac{\dot{\pi}^2}{ A}\right) 
  \right], \\
& \cG = 2 \phi \left[G_4 - 2 G_{4X} \left(X - \frac{\dot{\pi}^2}{2 A}\right)
+G_{5\pi} \left(X - \frac{\dot{\pi}^2}{ A}\right)
\right], \\
& \cH = 2 \phi \left[G_4 - 2 G_{4X} X + G_{5\pi} X \right] , \\
& \cJ = 2 \phi  \; \dot{\pi} \pi' ( G_{4X} -  G_{5\pi} ) .
\end{align} 
\end{subequations}
As for the $b_5$ coefficient in the action~\eqref{eq:action_odd_metric} 
let us note that it involves two contributions: the mass term $b_{5(m)} \cV^2$ and the angular part of the Laplace operator $ \ell(\ell+1) b_{5(\ell^2)} \cV^2$. While the mass term is irrelevant for computation of the vector's speed (so we omit the explicit form of $b_{5(m)}$ for brevity), the $ \ell(\ell+1)$  part defines the propagation speed of the photon in the angular direction, so we write it separately:
\[
\label{eq:NangularA}
b_{5(\ell^2)} = -\frac{\ell(\ell+1)}{2 J^4} \sqrt{\frac{A}{B}} \phi^2 (\cF+\cG-\cH).
\]
Note that the dilaton field $\phi$ enters all the coefficients in eqs.~\eqref{eq:HGF} and~\eqref{eq:NangularA} as an overall factor and, as we will shortly see, does not contribute to the propagation speeds similarly to the homogeneous case in~\cite{KK}.

The action~\eqref{eq:action_odd_metric} does not involve $\dot{h}_0$, hence, $h_0$ is manifestly non-dynamical. To avoid solving differential equation for $h_0$ we adopt an approach inspired by the trick used in pure (beyond) Horndeski theory in with a static inhomogeneous metric and a linearly-time dependent Galileon, see e.g.~\cite{Kobayashi:odd,Ogawa,Takahashi:2016dnv,Baez:2022rdz}. The trick has been also recently adopted in~\cite{MSV} for dynamical background~\eqref{eq:backgr_metric}. 
So as a first step we integrate by parts and rewrite the action~\eqref{eq:action_odd_metric} as follows:
\begin{multline}
  \label{eq:action_odd_square}
S^{(2)}_{h+\cV} = \int \mbox{d}t\:\mbox{d}r \left[ 
\left(a_1 - \frac{2}{J^2}\frac{d}{dr}\left[ a_3 {J'}{J} \right] \right) h_0^2
+ \left(a_2 - \frac{2}{J^2}\frac{d}{dt}\left[ a_3 {\dot{J}}{J} \right]\right) h_1^2 
+  \left(a_4 + 8 a_3 \frac{\dot{J}J'}{J^2}\right) h_1 h_0 
\right.\\\left. 
+ a_3 \left[{\dot h_1}- h_0' + 2 \left(\frac{J'}{J}h_0 -\frac{\dot{J}}{J}h_1\right)  \right]^2 
+ b_1 (\cV' h_0 - \dot{\cV} h_1 )
\right.\\\left. 
+ b_2 \dot{\cV}^2 
+ b_3 {\cV'}^2
+ b_4 \dot{\cV} \cV' 
+ \ell(\ell+1) b_{5(\ell^2)} \cV^2+ b_{5(m)} \cV^2
\right].
\end{multline}
Then we introduce an auxiliary field $Q$: 
\begin{multline}
  \label{eq:action_odd_Q}
S^{(2)}_{h+\cV} = \int \mbox{d}t\:\mbox{d}r \left[ 
\left(a_1 - \frac{2}{J^2}\frac{d}{dr}\left[ a_3 {J'}{J} \right] \right) h_0^2
+ \left(a_2 - \frac{2}{J^2}\frac{d}{dt}\left[ a_3 {\dot{J}}{J} \right]\right) h_1^2 
+  \left(a_4 + 8 a_3 \frac{\dot{J}J'}{J^2}\right) h_1 h_0 
\right.\\\left. 
+ a_3 \left( 2 Q \left[ {\dot h_1}- h_0' + 2 \left(\frac{J'}{J}h_0 -\frac{\dot{J}}{J}h_1\right)\right] - Q^2 \right) 
+ b_1 (\cV' h_0 - \dot{\cV} h_1 )
\right.\\\left. 
+ b_2 \dot{\cV}^2 
+ b_3 {\cV'}^2
+ b_4 \dot{\cV} \cV' 
+ \ell(\ell+1) b_{5(\ell^2)} \cV^2+ b_{5(m)} \cV^2
\right].
\end{multline}
Note that we immediately restore the initial form of action~\eqref{eq:action_odd_square} upon making use of the equation of motion for $Q$.
At this point integration by parts enables one to shift all derivatives on $Q$ in eq.~\eqref{eq:action_odd_Q}, thus, making both $h_0$ and $h_1$ non-dynamical variables. Then the variation of action w.r.t. $h_0$ and $h_1$ gives the following constraint equations, respectively:
\begin{subequations}
\label{eq:constraints}
\begin{align}
& 2 \left(a_1  - \frac{2}{J^2}\frac{d}{dr}\left[ a_3 {J'}{J} \right] \right) h_0 + \left(a_4 + 8 a_3 \frac{\dot{J}J'}{J^2}\right) h_1 
+ 4 a_3 \frac{J'}{J} Q + 2 \frac{d}{dr}\left[a_3 Q\right] + b_1 \cV'= 0, \\
& 2 \left(a_2 - \frac{2}{J^2}\frac{d}{dt}\left[ a_3 {\dot{J}}{J} \right] \right) h_1 + \left(a_4 + 8 a_3 \frac{\dot{J}J'}{J^2}\right) h_0 - 4 a_3 \frac{\dot{J}}{J} Q - 2 \frac{d}{dt}\left[a_3 Q\right] - b_1 \dot{\cV} = 0.
\end{align} 
\end{subequations}
Solving eqs.~\eqref{eq:constraints} for $h_0$ and $h_1$ in terms of 
$\dot{\cV}$, $\cV'$, $Q$ and its derivatives, and substituting the result into the action~\eqref{eq:action_odd_Q}, we arrive to the quadratic action for two degrees of freedom -- $Q$ and $\cV$:
\begin{multline}
\label{eq:action_odd_final}
S^{(2)}_{h+\cV} = \int \mbox{d}t\:\mbox{d}r\:\sqrt{\frac{B}{A}} J^2
\frac{\ell(\ell+1)}{2(\ell-1)(\ell+2)}
\left[ \frac{1}{A}\frac{\mathcal{F}\mathcal{H}^2}{\cZ } \dot{Q}^2 
- \frac{B \cdot \mathcal{G}\mathcal{H}^2}{ \cZ } (Q')^2 
+ 2 \frac{B}{A} \frac{\cJ\mathcal{H}^2}{\cZ  } Q'\dot{Q} 
-\frac{\ell(\ell+1)}{J^2}\cdot \mathcal{H} Q^2 
\right.\\ \left.
 +
\phi^2\left(\frac{A (\ell+2)(\ell-1)}{B J^2} 
+ \frac{\phi^2 \cH^2 }{\cZ}(\dot{A}_1 - A_0')^2 \right) \left[ \frac{1}{A}\cF \dot{\cV}^2 
-  B \cG \cdot {(\cV')}^2 + 2 \frac{B}{A} \cJ \dot{\cV} \cV'
\right]
\right.\\\left.
- \phi^2\frac{\ell(\ell+1)}{J^2} \left( \frac{A (\ell+2)(\ell-1)}{B J^2}
\frac{\cZ}{\cH}
\right)\cdot\cV^2
+ \frac{2 \phi^2 \cH^2 }{\cZ} (\dot{A}_1 - A_0')
\left[\frac{1}{A} \cF \dot{Q} \dot{\cV} -  B \cdot\cG Q' \cV' + \frac{B}{A}\cdot \cJ ( Q' \dot{\cV}+\dot{Q} \cV') \right] +
\dots \right] \;,
\end{multline}
where we have introduced a notation 
\footnote{The coefficient for $\ell(\ell+1)\;\cV^2$ in eq.~\eqref{eq:action_odd_final} in fact does not involve a denominator and explicitly reads for the theory in question as follows: 
$\cZ/\cH = \cF+\cG-\cH = \phi \;(2 G_4 - 2 G_{5\pi} X)$ (see eqs.~\eqref{eq:HGF} and~\eqref{eq:Z}). 
The chosen form of the coefficient in terms of $\cZ$ enables us in what follows to write the angular speeds $c_{\theta}^2$ in a concise form.}
\[
\label{eq:Z}
\cZ = \mathcal{G} \mathcal{F} + \frac{B}{A} \cdot\cJ^2.
\]
The ellipsis in eq.~\eqref{eq:action_odd_final} denotes the omitted terms $Q\dot{\cV}$, $Q\cV'$ and the mass terms for both $Q$ and $\cV$, which are all irrelevant in what follows since we are going to focus on deriving the propagation speeds for both $Q$ and $\cV$. 
Note that upon taking the limit $\phi \to \mbox{const}$ and $\cV \to 0$ in eq.~\eqref{eq:action_odd_final},
one can recover the result for pure linearized Horndeski theory over the background~\eqref{eq:backgr_metric}, see~\cite{MSV}.

\subsection{Dispersion relations and propagation speeds}
\label{sec:speeds}
Let us rewrite the quadratic action~\eqref{eq:action_odd_final} in a matrix form:
\[
\label{action_odd_matrix}
S^{(2)}_{h+\cV} = \int \mbox{d}t\:\mbox{d}r\:\sqrt{\frac{B}{A}} J^2
\frac{\ell(\ell+1)}{2(\ell-1)(\ell+2)}\cdot 
\left[ \mathcal{K}_{ij}\dot{v}^i\dot{v}^j - \mathcal{G}_{ij} {v^i}' {v^j}' + 
\mathcal{Q}_{ij}\dot{v}^i {v^j}' - \ell(\ell+1)\mathcal{M}_{ij(\ell^2) }v^i v^j+ \dots \right] \; ,
\]
where $v^1=Q$, $v^2=\cV$. 
The matrices $\cK_{ij}$, $\cG_{ij}$, $\cQ_{ij}$ and $\cM_{ij(\ell^2)}$ can be
straightforwardly read off from eq.~\eqref{eq:action_odd_final}.

We now focus on a high momentum regime, which implies that both $Q$ and $\cV$ vary with time and coordinate at much shorter scales than the background values in eqs.~\eqref{eq:backgr_metric} and~\eqref{eq:backgr_fields}. 
Then upon switching to the Fourier space and neglecting the sub-leading terms, we write
\[
\label{eq:dispersion}
 \left.\left( \omega^2  \mathcal{K}_{ij} - k^2 \mathcal{G}_{ij} - \omega k \mathcal{Q}_{ij} - \ell(\ell+1)\mathcal{M}_{ij(\ell^2)} \right)\right|_{{Eigenvalues}} = 0, 
\]
where the eigenvalues of the matrix in round brackets give the 
dispersion relations for $Q$ and $\cV$. 
Then the speeds of propagation in the radial direction are given by the solutions of the following equations
\footnote{This form of equation implies that one finds the eigenvalues of the matrix in brackets and then equates it to zero, so that the solution of the resulting equation gives the propagation speed.}:
\[
\label{eq:eigenR}
\left. \left( c_r^2 \cdot \mathbf{I}_{ij} - (AB)^{-1} \cdot \mathcal{K}_{ik}^{-1}\mathcal{G}_{kj} - c_r \cdot (AB)^{-1/2} \cdot \mathcal{K}_{ik}^{-1}\mathcal{Q}_{kj}\right)\right|_{{Eigenvalues}} = 0.  
\]
We find that the eigenvalues of eq.~\eqref{eq:eigenR} are identical and explicitly read:
\[
\label{eq:speed_radial_dispersion}
c_r^2 - 2 \sqrt{\frac{B}{A}}\frac{ \cJ }{\cF} \cdot c_r - \frac{\cG}{\cF}= 0,
\]
hence, the radial speeds are:
\[
\label{eq:speed_radial}
c_r^{(\pm)} = \sqrt{\frac{B}{A}}\frac{\cJ}{\cF} \pm \frac{1}{\cF}\sqrt{ \mathcal{Z}}.
\]
Thus, both odd-parity modes $Q$ and $\cV$, which can be identified with the graviton and the photon respectively, have matching inward and outward propagations speeds in the radial direction:
\[
\label{eq:radial_equality}
c^{+}_{r, Q} = c^{+}_{r, \cV}, \quad c^{-}_{r, Q} = c^{-}_{r, \cV}.
\]

As for the propagation speeds in the angular direction, they can be found 
from the following relations
\[
\label{eq:eigenA}
\left( c_{\theta}^2 \cdot \mathbf{I}_{ij} - \left.\frac{J^2}{A} \mathcal{K}_{ik}^{-1}\mathcal{M}_{kj}\right) \right|_{Eigenvalues} = 0.
\]
Again we find that the eigenvalues in eq.~\eqref{eq:eigenA}
coincide and so do the angular speeds for both the graviton $Q$ and the photon $\cV$. The speeds explicitly read:
\[
\label{eq:speed_angular}
c_{\theta, Q}^2= c_{\theta, \cV}^2 = \frac{\cZ}{\cF\cH}.  
\]
{Let us note that generally both $c_{r,Q}^{(\pm)}=c_{r,\cV}^{(\pm)}$
and $c_{\theta,Q}^2=c_{\theta,\cV}^2$ are non-unit except for a deliberate choice of Lagrangian functions in~\eqref{eq:G4_5DGalileonsIn4D} and~\eqref{eq:G5_5DGalileonsIn4D}.
}

Note that the results for propagation speeds for
the gravity mode in eqs.~\eqref{eq:speed_radial} and~\eqref{eq:speed_angular} are in agreement with the results in pure (beyond) Horndeski theory, see~\cite{MSV}.
Moreover, the original results of~\cite{KK} in a cosmological background can be also restored. For the homogeneous case one has $X \to \dot{\pi}^2/2$ 
 and, according to eqs.~\eqref{eq:HGF}, $\cJ \to 0$ and adopting the notations of~\cite{KK} $\cF = \cH \equiv \cG_{\mathcal{T}}$, $\cG \equiv \cF_{\mathcal{T}}$, so that both the speed of a tensor mode and that of light  are equal and read:
\[
c^2_{\mathcal{T}}=c^2 = \frac{2 G_4}{2G_4 - 4 G_{4X}X +2 G_{5\pi}X}.
\] 
Similarly one can derive the corresponding speeds for the case of a static, spherically symmetric background with $\cJ \to 0$ and $\dot{\pi} \to 0$ with $X \to -B{\pi'}^2/2$ in eqs.~\eqref{eq:HGF} (c.f.~\cite{Kobayashi:odd}):
\[
c_{r,Q}^{2}=c_{r,\cV}^{2} = \frac{\cG}{\cF} = \frac{G_4 - 2 G_{4X}X +G_{5\pi} X }{G_4 -G_{5\pi}X},
\qquad
c_{\theta,Q}^2=c_{\theta,\cV}^2 = \frac{\cG}{\cH} = 1.
\]

\subsection{Stability constraints for high energy modes with $\ell \geq 2$}
\label{sec:stability}

Let us briefly address the set of stability criteria, which ensure the absence of ghosts and gradient instabilities among the high energy modes. 

The constraints immediately follow from eq.~\eqref{eq:dispersion} and eqs.~\eqref{eq:speed_radial} and~\eqref{eq:speed_angular}.
Namely, one has to ensure that the kinetic matrix $\cK_{ij}$ is positive definite to avoid ghosty modes:
\[
\label{eq:K11detK}
\cK_{11} = \frac{1}{A}\frac{\mathcal{F}\mathcal{H}^2}{\cZ } > 1, 
\quad 
\det\cK = \phi^2\frac{(\ell-1)(\ell+2)}{B J^2} \frac{\cF^2\cH^2}{\cZ} > 0.
\]
Thus, the no-ghost condition explicitly reads:
\[
\label{eq:no-ghost}
\cF >0, \qquad \cZ = \mathcal{G} \mathcal{F} + \frac{B}{A} \cdot\cJ^2 >0.
\]
Note that the positivity of $\cZ$ also ensures that the radial speeds~\eqref{eq:speed_radial} are real and, hence, there are no radial gradient instability.
As for the angular propagation speed~\eqref{eq:speed_angular}, it gives the 
following criteria for angular gradient instabilities to be absent:
\[
\label{eq:no_gradient_angular}
\cH > 0.
\]
The resulting set of stability constraints~\eqref{eq:no-ghost} and~\eqref{eq:no_gradient_angular} is identical to that obtained for pure (beyond) Horndeski theory in~\cite{MSV}.
Let us note that we opt for strict inequalities in the constraints above in order to be on a safe side and avoid potential problems with strong coupling or degeneracy in dispersion relations.

\section{Discussion and conclusions}
\label{sec:conclusion}

In this work we have revisited the implications of multi-messenger speed tests for Horndeski theories as the candidates for late-time modifications of gravity in a homogeneous Universe.  
In particular, the existing constraints from GW170817 event force one to ensure that the speed of GW is utterly close to that of light. 
Taking this limitation to the extreme and requiring strict equality $c_{GW} = c_{\gamma}$, one faces two possible approaches to comply with this equality within Horndeski theories on theoretical grounds.

First, one can assume that the photon is minimally coupled and, hence,
$c_{GW} = c_{\gamma}=1$. In this case Horndeski theories were shown to be severely constrained, provided one avoids fine-tuning of the background, so that leftover models are relatively simple, 
i.e. with $G_{4X} = 0$ and $G_5 = \mbox{const}$.
If one allows beyond Horndeski generalization~\cite{Zumalacarregui:2013pma,Gleyzes:2014dya},
there is a certain loosening of constraints and models with $G_{4X} \neq 0$ and $F_{4} = G_{4X}/2X$ are also compliant with the speeds' equality in a cosmological setting~\cite{Ezquiaga:2017ekz}. 

Second, as an alternative one may allow non-trivial couplings for a photon, which generally results in $c_{\gamma} \neq 1$. 
In particular, this paper has addressed the relation between $c_{GW}$ and $c_{\gamma}$ within a subclass of Horndeski theory with non-minimally coupled vector field, formulated earlier in~\cite{KK}.
The key finding of~\cite{KK} is that this class of scalar-vector-tensor theories features both GW and electromagnetic waves propagating with the same speed in a cosmological setting for an arbitrary choice of $G_4(\pi,X)\neq0$ and $G_5(\pi)\neq0$. The latter significantly relaxes the original constraints of GW170817 and revives a wide range of Horndeski subclasses, at least from the multi-messenger speed test point of view. 

In this paper we have generalized the result of~\cite{KK} and proved that the equality $c_{GW} = c_{\gamma}$ holds even upon departures from the homogeneous FLRW background. 
Namely, we have derived the propagation speeds in both radial and angular directions for the graviton and the photon above the time-dependent, spherically symmetric background of a general form and showed that the speeds are identical for arbitrary scalar potentials $G_4(\pi,X)$ and $G_5(\pi)$ in the Lagrangians~\eqref{eq:G4_5DGalileonsIn4D} and~\eqref{eq:G5_5DGalileonsIn4D} without any fine-tuning of the background. 
Hence, this result supports the claim that a considerable number of Horndeski subclasses get recovered from GW170817 stop-list if a non-minimal Galileon-style coupling of a photon is allowed.

We have also derived a set of constraints for Lagrangian functions which ensure the absence of pathological degrees of freedom like ghosts and gradient instabilities among the high energy perturbation modes:
$$ \cF >0, \qquad \cZ = \cG\cH + \frac{B}{A}\cdot \cJ^2 > 0,
\qquad \cH > 0.$$

Let us note in conclusion that the scalar-vector-tensor theories~\eqref{eq:G4_5DGalileonsIn4D}-\eqref{eq:G5_5DGalileonsIn4D} are not unique in a sense that there may be other forms of photon couplings, which result in equality between the graviton and photon speeds 
(see e.g.~\cite{Babichev:2024kfo}). However,
the advantage of the current theory is that 
it arises naturally in result of Kaluza-Klein reduction of a pure 5D Horndeski theory, where a specific form of Galileon-vector couplings, involved in the resulting 4D Lagrangian, are similar to Galileon-gravity terms. Thanks to this form of non-trivial couplings  
the gravitational and electromagnetic waves propagate with the same, although generally non-unit, speed.
In view of higher dimensional origin of the theory the equality of the graviton's and photon's propagation speeds in not particularly surprising: 
it appears only natural since in the context of compactification the GW and the electromagnetic wave in 4D descend from the sole 5D GW. 
On the other hand, 
an elegant trick of dimensional reduction may be seen just as a useful tool for obtaining a particular form of vector-scalar couplings which in the end are compliant with the equality $c_{GW} = c_{\gamma}$.

\section*{Acknowledgements}
The authors are grateful to M. Valencia-Villegas and A. Shtennikova for 
valuable discussions.
The work on this project
has been supported by Russian Science Foundation grant № 24-72-10110,
\href{https://rscf.ru/project/24-72-10110/}{  https://rscf.ru/project/24-72-10110/}.


\end{document}